\documentclass[aps,pra,twocolumn,groupedaddress]{revtex4}

\usepackage[latin1]{inputenc}
\usepackage{amsmath}
\usepackage{graphicx}
\usepackage{amsfonts}

\newcommand*{\+}{\dagger}
\newcommand*{\bra}[1]{\mathopen{\langle}#1\mathclose{|}}%
\newcommand*{\brao}[1]{\bra{#1}}%
\newcommand{\braket}[2]{\mathopen{\langle}#1\mathclose{|}#2\mathclose{\rangle}}%
\newcommand{\braketo}[2]{\mathopen{\langle}#1 \mathclose{|}#2 \mathclose{\rangle}}%
\newcommand*{\Cop}[1]{\op{C}_{#1}}%
\newcommand*{\dbar}{\mkern 3mu\mathchar '26\mkern -12mu\dd}%
\newcommand{\dd}{\mathrm{d}}%
\newcommand*{\Domseff}{\Delta \om{s}^\mathrm{eff}}%
\newcommand*{\dotHeff}[1]{\dot{\op H}^\mathrm{eff}_{#1}}%
\newcommand*{\ee}{\mathrm{e}}%
\newcommand*{\expv}[1]{\mathopen{\langle}#1\mathclose{\rangle}}%
\newcommand*{\Heff}[1]{\op H^\mathrm{eff}_{#1}}%
\newcommand*{\Hop}[1]{\op H_{#1}}%
\newcommand*{\Hopm}[1]{\op H_\mathrm{#1}}%
\renewcommand*{\Im}{\mathrm{Im}}%
\newcommand*{\ket}[1]{\mathopen{|}#1\mathclose{\rangle}}%
\newcommand{\ketbra}[2]{\mathopen{|}#1\mathclose{\rangle\langle}#2\mathclose{|}}%
\newcommand*{\keto}[1]{\ket{#1}}%
\newcommand*{\kets}[1]{\ket{#1}}%
\newcommand*{\Leff}[1]{\mathcal{L}^\mathrm{eff}_{#1}}%
\newcommand*{\om}[1]{\omega_{#1}}%
\newcommand*{\op}{\hat}%
\renewcommand*{\Re}{\mathrm{Re}}%
\newcommand*{\rop}[1]{\op{\rho}_{#1}}%
\newcommand*{\tr}{\mathrm{tr}}%

\begin{document}


\title{Work exchange between quantum systems: the spin-oscillator model}


\author{Heiko Schröder}
\email[]{heiko.schroeder@itp1.uni-stuttgart.de}
\author{Günter Mahler}
\email[]{guenter.mahler@itp1.uni-stuttgart.de}
\affiliation{Institut für Theoretische Physik 1, Universität Stuttgart, Pfaffenwaldring 57, 70550 Stuttgart, Germany}


\date{\today}

\begin{abstract}
  With the development of quantum thermodynamics it has been shown
  that relaxation to thermal equilibrium and with it the concept of
  heat flux may emerge directly from quantum mechanics. This happens
  for a large class of quantum systems if embedded into another
  quantum environment. In this paper, we discuss the complementary
  question of the emergence of work flux from quantum mechanics. We
  introduce and discuss two different methods to assess the work
  source quality of a system, one based on the generalized
  factorization approximation, the other based on generalized
  definitions of work and heat. By means of those methods, we show
  that small quantum systems can, indeed, act as work reservoirs. We
  illustrate this behavior for a simple system consisting of a spin
  coupled to an oscillator and investigate the effects of two
  different interactions on the work source quality. One case will be
  shown to allow for a work source functionality of arbitrarily high
  quality.
\end{abstract}


\maketitle



\section{Introduction}
\label{sec:in}

Thermodynamics is a theory of impressive success and a wide range of
applicability. This is true despite its origin as a purely
phenomenological theory in the late 16th and 17th century. It took
about two hundred years until Boltzmann and Maxwell gave a foundation
of thermodynamics in terms of statistical mechanics based on classical
mechanics. It was only in recent years that a new approach to the
foundation of thermodynamics from a microscopic theory has received
increased attention: quantum thermodynamics, the derivation of
thermodynamics from quantum theory. In the seminal work of Gemmer et
al. \cite{gemmer04,BGM03b,GOM01}, the emergence of thermodynamic
behavior from quantum mechanics for a wide class of quantum systems
has been established. Other recent papers have extended and clarified
the results of the aforementioned authors
\cite{PSW06,LPS+08,goldstein:050403}.

The interest in a quantum approach to thermodynamics is two-fold:
First, it promises a deeper understanding of thermodynamic core
concepts like relaxation, irreversibility, heat, work, and might
elucidate connections between those and concepts known from quantum
mechanics, e.g., entanglement. Second, such an approach should help to
find generalizations for the mentioned concepts for non-equilibrium,
finite systems and strong interaction.

The definition of work in quantum systems has been discussed in
various papers \cite{Ali79,KR84,TLH07} and have been applied
successfully to quantum heat engines
\cite{Kie04,HMM06,HM07,HRM07,allahverdyan:041118}. Still, all those
investigations typically deal with quantum systems that are subject to
driving by means of a time-dependent Hamilton operator of the
system. Thus, the identification and definition of work is determined
\emph{a priori} by relating it to the presence of classical driving,
while no microscopic derivation of the concept is given. In the
present paper we deal with closed, finite quantum systems. For such
systems the functionality of cooling has already been investigated
(see, e.g. \cite{LPS}). Here we are interested in the question under
what conditions a quantum system coupled to another can exert the
effect of a classical driver over the other system and thus be
identified with a reversible work source. By our new approach based on
a complete quantum modelling of the work source, we are able to show
that classical driving and therefore work is not a concept bound to
macroscopic devices. 

The paper is organized as follows. In Sec.~\ref{sec:effdyn}, we
present the factorization approximation (FA) and its generalization to
the case of semi-mixed factorizing initial states. It is shown that
the applicability of the FA allows one to identify quantum systems as
classical drivers. In the subsequent Sec.~\ref{sec:q-meas}, we deal
with the question of work source quality definition. Based on the
previous section, we introduce a measure inspired by the FA and
establish its connection to work source functionality. In addition, we
develop another work reservoir quality measure based on considerations
of work and heat fluxes, if an appropriate definition of those is
given. The definition of work and heat flux we chose to use throughout
the paper is taken from \cite{WHRSM08} and outlined in
Sec.~\ref{sec:lembas}.  In Sec.~\ref{sec:model} we present the
spin-oscillator model and its properties for two types of
interactions. This simple quantum model is then used to illustrate the
implementation of work sources of arbitrarily high quality in quantum
mechanics in Sec.~\ref{sec:z-work} and to discuss the limits of two
different work functionality quality measures in
Sec.~\ref{sec:xz-work}, where we also give an idea of the overall work
source behavior of the second type of the model. Finally, we summarize
our results in Sec.~\ref{sec:sum}.

\section{Effective Dynamics: Time-dependent Driving}
\label{sec:effdyn}

The factorization approximation (FA) has been thoroughly discussed,
e.g., in \cite{gemmer0108} and \cite{WP74}. We will therefore only
summarize the basic statements and give a generalization to the result
of \cite{gemmer0108}.

In its form stated in \cite{gemmer0108}, the FA reads as follows. Let
us consider a bipartite quantum system with Hamilton operator
\begin{equation}
  \label{eq:1}
  \op H = \Hop{1} + \Hop{12} + \Hop{2}
\end{equation}
acting on the joint Hilbert space
$\mathcal{H}_{12}=\mathcal{H}_1\otimes\mathcal {H}_2$. The operators
$\Hop{1}$ and $\Hop{2}$ act on the respective local Hilbert spaces
$\mathcal{H}_1$ and $\mathcal{H}_2$ only. Let the initial state
factorize, i.e.,
\begin{equation}
  \label{eq:2}
  \ket{\Psi(0)} = \ket{\Psi_1(0)}\otimes\ket{\Psi_2(0)}
  \, .
\end{equation}
After some time $t$, the total state of the system either given by
$\ket{\Psi(t)}$ or its density matrix
$\op\rho(t)=\ketbra{\Psi(t)}{\Psi(t)}$ gives rise to the reduced
states of the two subsystems
$\op\rho_1(t)=\tr_2(\ketbra{\Psi(t)}{\Psi(t)})$ and
$\op\rho_2(t)=\tr_1(\ketbra{\Psi(t)}{\Psi(t)})$. Although the state
was assumed to factorize initially, in general, the subsystem states
no longer will be pure states due to entanglement introduced by the
interaction between the subsystems. As the total state is pure, the
subsystem purities $P[\op\rho_1(t)]=\tr\{[\op\rho_1(t)]^2\}$ and
$P[\op\rho_2(t)]=\tr\{[\op\rho_2(t)]^2\}$ are equal at any instant
$t$. Now, as long as the purity of the subsystems is close to 1, it
can be shown that the dynamics of the system are, in good
approximation, given by the reduced density matrices $\op\rho_i =
\ketbra{\Psi_i(t)}{\Psi_i(t)}$, $i=1,2$, where the $\ket{\Psi_i(t)}$
obey the coupled differential equations
\begin{eqnarray}
  \label{eq:3}
  i \hbar \ket{\dot\Psi_1(t)} &=& \left(\Hop{1}+\bra{\Psi_2(t)}\Hop{12}\ket{\Psi_2(t)}\right)\ket{\Psi_1(t)}\\
  \label{eq:4}
  i \hbar \ket{\dot\Psi_2(t)} &=& \left(\Hop{2}+\bra{\Psi_1(t)}\Hop{12}\ket{\Psi_1(t)}\right)\ket{\Psi_2(t)}
\end{eqnarray}
up to an irrelevant relative phase (for a detailed derivation see
\cite{gemmer0108}). Obviously, the reduced states of the system evolve
under the action of time-dependent effective Hamiltonians
$\Hop{1}+\Heff{1}(t)$ and $\Hop{2}+\Heff{2}(t)$, respectively, with
\begin{eqnarray}
  \label{eq:5}
  \Heff{1}(t)&=&\tr_2\{\Hop{12}[\op1\otimes\op\rho_2(t)]\} \\
  \label{eq:6}
  \Heff{2}(t)&=&\tr_1\{\Hop{12}[\op\rho_1(t)\otimes\op 1]\} 
\end{eqnarray}
and in the present case $\rop{j}(t) = \ketbra{\Psi_j(t)}{\Psi_j(t)},
j=1,2$.

The above statement can be generalized for the case of factorizing
semi-mixed initial states, that is, states of the form
\begin{equation}
  \label{eq:7}
  \op\rho(0)=\op\rho_1(0)\otimes\ketbra{\Psi_2(0)}{\Psi_2(0)}
  \, .
\end{equation}
If the purity $P[\op\rho_2(t)]$ of the initially pure system 2 remains
close to unity, the dynamics of the system can be given approximately
by the coupled differential equations
\begin{eqnarray}
  \label{eq:genFAmixed}
  i \hbar \dot{\op{\rho}}_1(t) &=& \left[\Hop{1}+\Heff{1}(t),\op\rho_1(t)\right]\\
  \label{eq:genFApure}
  i \hbar \ket{\dot{\Psi}_2(t)} &=& \left(\Hop{2}+\Heff{2}(t)\right)\ket{\Psi_2(t)}
\end{eqnarray}
with $\op\rho_2(t)=\ketbra{\Psi_2(t)}{\Psi_2(t)}$.  Again, the effect
of the subsystems on each other is to induce a time-dependent
effective Hamiltonian that governs the time evolution of the
subsystems. For the derivation, see App.~\ref{app:gfa}.

Here, we would like to stress the fact that, as long as the
prerequisites for the FA are met,
Eqs.~(\ref{eq:genFAmixed},~\ref{eq:genFApure}) present an alternative
description of the system dynamics: The subsystems can be considered
classical drivers for each other. It is remarkable that this feature
is reciprocal and based on (approximate) constancy of the subsystem
purities (entropies). 

The energy exchanged this way can aptly be called ``work'':
Classically one would define the work $W$ imparted over time $t_S$ on
a Hamiltonian system $H(\lambda)$ with $\lambda$ denoting the
time-dependent control parameter as \cite{PhysRevE.56.5018}
\begin{equation}
  \label{eq:8}
  W = \int\limits_0^{t_S} \dd t \frac{\dd\lambda}{\dd t}\frac{\partial H}{\partial \lambda}[\vec{z}(t)],
\end{equation}
where $\vec{z}(t)$ denotes the system's state trajectory in phase
space. One notes, however, that the energy exchange will, in general,
be contaminated by contributions violating the constancy of local
purity. This contamination is a characteristic feature of the
underlying total (unitary) dynamics. Close to thermal equilibrium such
a contribution would be called heat, $\dbar Q$: Work and heat in open
quantum systems are usually defined as
\cite{Ali79,KR84,Kie04,HMM06,HM07}
\begin{equation}
  \label{eq:9}
  \dd U = \dd \expv{\Hop{}} =
  \underbrace{\tr(\op \rho \dd \Hop{})}_{\dbar W} + \underbrace{\tr(\Hop{} \dd \op \rho)}_{\dbar Q}
\end{equation}
again recognizing the energy exchange in the FA scenario as work.

We emphasize here, that explicitly time-dependent Hamiltonians are not
part of the fundamental description of nature as given by quantum
mechanics. Therefore, there is no way how they could come about save
by an effective description of a system like the FA. If one denied any
physical significance of such an effective description and hence
considered it only a mathematical simplification without physical
meaning, one obviously would have to deny the physical existence of
classical drivers altogether. This is not a reasonable option. 

\section{LEMBAS principle}
\label{sec:lembas}

The effective dynamics according to
Eqs.~(\ref{eq:genFAmixed},\ref{eq:genFApure}) allows for an intuitive
approach to the concept of work: In general, however, only
approximately; the deviations remain unquantified.

Here, the LEMBAS approach \cite{WHRSM08} comes into play based on the
following ideas: First, choose a partitioning of the total isolated
system into system of interest (1) and its environment (2) and
consider the exact local dynamics of the system (1). The state of the
total system is
\begin{equation}
  \label{eq:10}
  \rop{} = \rop{1}\otimes\rop{2} + \op C_{12}
\end{equation}
where $\rop{j}$ are the respective reduced density operators. Then,
the exact (effective) Liouville-von Neumann equation for subsystem (1)
can be written as
\begin{equation}
  \label{eq:11}
  \dot\rho_{1}(t) = -\frac{i}{\hbar} \left[\Hop{1}+\Heff{1}(t),\op\rho_1(t)\right] +\Leff{1}[\op\rho(t)]
\end{equation}
with the superoperator
$\Leff{1}[\op\rho(t)]=-i\hbar^{-1}\tr_2\{[\Hop{12},\Cop{12}]\}$.

The local energy $\Hop{1}'$ is defined now based on considerations how
the local system would appear to an experimenter (``local effective
measurement basis'', LEMBAS). There is some ambiguity in the
procedure, but it has proven useful in \cite{WHRSM08} to choose
\begin{equation}
  \label{eq:12}
  \Hop{1}' = \Hop{1} + \Heff{1,a}(t)
\end{equation}
where $\Heff{1}(t) = \Heff{1,a}(t) + \Heff{1,b}(t)$ and
$\Heff{1,a}(t)$ is the part of $\Heff{1}(t)$ that commutes with
$\Hop{1}$.

The final step is to discriminate energy changes of the system based
on whether they change the local von Neumann entropy $S_1$ or not,
that is whether they are of coherent (work) or incoherent origin
(heat). This leads to the following formulas for heat- and work-flux
for any partitioning and any $\Hop{1}'$:
\begin{eqnarray}
  \label{eq:13}
  \dot W(t) &=& \tr\{\dotHeff{1,a}(t) \rop{1}(t) - i [\Hop{1}'(t),\Heff{1,b}(t)] \rop{1}(t) \} \\
  \label{eq:14}
  \dot Q(t) &=& \tr\{\Hop{1}'(t)\Leff{1}[\rop{}(t)]\}
  .
\end{eqnarray}

How do these generalized definitions connect to their thermodynamic
analogues? In the thermodynamic limit, that is, close to the
thermodynamic equilibrium, for infinitely sized systems and weak
couplings, the von Neumann entropy of the respective subsystem and its
thermodynamic entropy coincide and the LEMBAS definitions of work and
heat blend in with their thermodynamic counterparts. 

But also in far from equilibrium situations, the LEMBAS definitions
can be associated with work and heat in the following sense: We know
from the results of quantum thermodynamics
\cite{gemmer04,BGM03b,GOM01} that thermodynamic behavior of a system
can be seen to result from an embedding in an environment, which by
itself needs not to be and usually is not thermodynamic (in
equilibrium, infinite, weak coupling). Thus, validity of thermodynamic
concepts is not a property of the total system but has to do with
whether or not the system of interest is influenced by its environment
in such a way that thermodynamic properties emerge, which is a purely
local consideration. The LEMBAS definitions take this concept to the
extreme in the sense that they state that ``what locally has a work
effect $\Heff{1}(t)$, is work'' and ``what locally has a heat effect
$\Leff{1}[\rop(t)]$, is heat'' even for non-thermodynamic (in the
classical sense), far from equilibrium situations. Making the
distinction in this way is justified by the fact that classical
driving can be unambiguously identified as work even in the
thermodynamic sense and, therefore, any effect $\Leff{1}[\rop(t)]$ not
related to work is identified as heat.

Finally, we note that the LEMBAS definitions retain the properties
that
\begin{enumerate}
\item work is energy exchange due to changing parameters of the
  Hamilton operator that describes the system;
\item heat is energy exchange associated with change of entropy,
  although here a generalized definition of entropy is to be used.
\end{enumerate}

\section{Measures of work source quality}
\label{sec:q-meas}

\subsection{Work reservoir}
\label{sec:q-meas:wsource}

An ideal work reservoir can be defined as a system exchanging energy
only in the form of work. It is obvious that this definition is too
restrictive for the classification of realistic models, that is,
models involving finite size, finite interaction and limited
control. No realistic model can comply to the idea of such an ideal
work source as even arbitrary small but finite deviations from this
idealized concept would lead to a rejection of a model as a work
source. Additional complications arise due to the fact that we have to
consider processes, the properties of which may change with time.

Thus, there is need for a more differentiated measure of work
reservoir functionality. In a non ideal world, special attention is to
be paid to the definition and quantification of the quality of a work
reservoir to be able to compare and to draw conclusions on justified
grounds.

Basically, one can distinguish two types of measures depending on
whether they refer to a single point in time or to a (finite or
infinitely large) interval of time. We like to refer to them as
\emph{instantaneous} and \emph{integral} measures and our main
interest lies on the integral ones, defined with respect to some
finite time interval (again because under realistic condition it is
not expected that a system can be a work source for all times).

In the following section, we present two different approaches to the
problem based on two distinct physical reasonings.

\subsection{Purity based measure}
\label{sec:q-meas:p}

Comparing Eq.~(\ref{eq:11}) to Eq.~(\ref{eq:genFAmixed}), one realizes
that the applicability of the FA is equivalent to a vanishing
$\Leff{1}$. Thus, if the total system was initially in a semi-mixed
state, $\Leff{1}$ is negligible if $P[\rop{2}(t)]\approx1$. In this
sense, $P[\rop{2}(t)]$ is a measure of work reservoir
functionality. The closer it is to 1, the smaller $\Leff{1}$ has to be
and the less energy may be exchanged as heat instead of work. Note,
that acting as a work reservoir is a reciprocal property, i.e., each
subsystem acts on its partner in an analogous way. This is in perfect
agreement with what we know from thermodynamics. If we have two
systems undergoing a process during which only work is exchanged
between them, both systems obviously act as work reservoirs for each
other although we may imagine one system to be the gas filling a box
and the other system to be the piston capping the box and being
connected to a spring.

At first glance, the purity therefore seems to be a good candidate for
assessing work source quality: It is an easy quantity to compute --
even analytically -- and by its connection to the FA, the physical
reasoning is clear.

However, as clear as the ideal situation with $P[\rop{2}(t)]=1$ is, it
is unclear to give a quantitative interpretation for purities lower
than unity because there is neither an obvious relation between
$P[\rop{2}(t)]$ and $\Leff{1}$ nor between $\Leff{1}$ and the quality
of the work reservoir functionality. Moreover, it is expected that the
same purity decrease for different systems, especially of different
size, has to be weighted differently. Thus, any concrete choice of a
minimum purity beyond which a system will be accepted as a work
reservoir will remain somewhat arbitrary and difficult to compare with
other systems' purity behavior. If such a purity threshold was given,
the respective system could be considered as a work source for any
time interval during which the purity stays above the given threshold.

As will become evident in Sec.~\ref{sec:model:xzint}, there is another
problem besides the arbitrary definition of the threshold when using
this measure: The decrease in purity is linked to the size of
$\Leff{1}$ only. Thus, the purity does not contain any information
about the relative effects of $\Heff{1}(t)$ and $\Leff{1}$. Since the
former is related to work and the latter to heat, a comparison of both
in terms of their effect on the energy of the system is in general
expected to be an important part of the assessment of work source
quality.

\subsection{Work and heat flux based measure}
\label{sec:q-meas:wq}

We introduce the ratio
\begin{equation}
  \label{eq:15}
  r(t) := \frac{|\dot W(t)|}{|\dot W(t)|+|\dot Q(t)|}
\end{equation}
which has the following convenient properties:
\begin{itemize}
\item $r(t) = 1 \Leftrightarrow \dot W(t) \neq 0 \land \dot Q(t) = 0$:
  ideal work source
\item $r(t) = 0 \Leftrightarrow \dot W(t) = 0 \land \dot Q(t) \neq 0$:
  ideal heat source
\end{itemize}
Provided there is energy exchange at all (i.e. not both, $\dot W$,
$\dot Q$ are zero), $r$ is well behaved. As we took separate moduli in
the denominator, there can be no compensation due to opposite sign.

Based on this instantaneous measure, we can develop an integral
measure for finite time intervals $[t_0,t_1]$. Directly integrating
over $r(t)$ is not an option for this would completely ignore the
time-dependence of the total of the absolute fluxes and therefore the
necessary weighting of $r$. It is straightforward to apply the
necessary weight, integrate and then normalize the result defining
\begin{eqnarray}
  R(t_1,t_0) 
  &:=&
  \frac{\int\limits_{t_0}^{t_1} r(t) \left( |\dot W(t)|+|\dot Q(t)| \right) \dd t}{\int\limits_{t_0}^{t_1} \left( |\dot W(t)|+|\dot Q(t)| \right) \dd t}
  \nonumber \\
  \label{eq:16}
  &=&
  \frac{\int\limits_{t_0}^{t_1} |\dot W(t)| \dd t}{\int\limits_{t_0}^{t_1} \left( |\dot W(t)|+|\dot Q(t)| \right) \dd t}
  .
\end{eqnarray}
Defining the quantities
\begin{equation}
  \label{eq:17}
  \mathcal{W}(t_1,t_0) := \int\limits_{t_0}^{t_1} |\dot W(t)| \dd t
  ,\,
  \mathcal{Q}(t_1,t_0) := \int\limits_{t_0}^{t_1} |\dot Q(t)| \dd t
\end{equation}
we can rewrite Eq.~(\ref{eq:16}) in the form of Eq.~(\ref{eq:15}) as
\begin{equation}
  \label{eq:18}
  R(t_1,t_0) := \frac{\mathcal{W}(t_1,t_0)}{\mathcal{W}(t_1,t_0)+\mathcal{Q}(t_1,t_0)} .
\end{equation}
This integral measure has the same special points like the
instantaneous measure with the following interpretations:
\begin{itemize}
\item $R(t) = 1 \Leftrightarrow \dot Q(t) = 0$ for all $t \in
  [t_0,t_1]$ and $\dot W(t)\neq0$ for some $t \in [t_0,t_1]$: ideal
  work source
\item $R(t) = 0 \Leftrightarrow \dot W(t) = 0$ for all $t \in
  [t_0,t_1]$ and $\dot Q(t) \neq 0$ for some $t \in [t_0,t_1]$: ideal
  heat source
\end{itemize}
We stress the fact here that a measure based on the integrated work
$W(t_1,t_0)=\int_{t_0}^{t_1} \dot W(t) \dd t$ and the (analogously)
integrated heat is not able to accomplish such precise assessment of
the work source quality: For oscillating fluxes, e.g., $Q(t_1,t_0)$
might reach 0 for some interval, although during the time interval
there might have flown vast amounts of heat. By employing the
integrals of the absolute fluxes in the chosen definition, we achieve
a much stronger statement about the quality of a system.

Finally, let us note that there is also a drawback to this measure,
namely the difficulty of calculating it because of the integration
over the absolute values of the fluxes.

\section{Application: Spin-Oscillator Model (SOM)}
\label{sec:model}

We turn now to the description of the model we will use to demonstrate
the existence of small quantum systems that do act as work sources. We
illustrate the features of FA and the various work measures we have
discussed above and discuss the model and its properties with special
focus on the dynamics of the purity.

The model is a single spin interacting with a harmonic oscillator
(\emph{spin-oscillator model, SOM}). On the one hand, the SOM serves as an
allusion to a classical steam engine with a gas of some temperature
(spin) and a piston periodically compressing and expanding the gas
(oscillator). On the other hand, the SOM has been used in previous
related works as a central element of quantum thermodynamic machines
\cite{TM06,HMM06,HM07}. Also, the simplicity and therefore partially
possible analytical treatment of the model has further motivated the
choice.

The SOM is defined by the Hamiltonian
\begin{equation}
  \label{eq:Hsom}
  \Hopm{}
  =
  \frac{\om{s}}{2} \op \sigma_z + \Hopm{int} + \om{o}\left(\op a^\dagger \op a + \frac12\right)
  ,
\end{equation}
where we have set $\hbar=1$. We denote the spin and oscillator local
Hamilton operators as $\Hop{s}$ and $\Hop{o}$, respectively.  The
eigenstates of $\Hop{s}$ are $\kets{0}$ and $\kets{1}$ with the
respective eigenvalues $\mp\om{s}/2$. The eigenstates of $\Hop{o}$ are
defined as $\{\keto{k}\}$ with eigenvalues $\om{o}(k+1/2)$, where
$k=0,1,2,\dots$.

We will discuss the $z$- and the $xz$-interaction,
$\Hopm{int}=\{\Hop{z},\Hop{xz}\}$, where
\begin{eqnarray}
  \label{eq:19}
  \Hopm{int} & = & \Hop{z} = \lambda \op\sigma_z\op x \\
  \label{eq:20}
  \Hopm{int} & = & \Hop{xz} = \lambda (\op\sigma_z + \kappa \op\sigma_x) \op x 
  .
\end{eqnarray}

For the initial state of the total system, we assume that the spin has
interacted in the past with some heat bath in order to establish a
thermal state but now is decoupled from said bath (or the bath
coupling is so weak that its influence may be neglected during the
period of evolution one is interested in). The oscillator is prepared
in a coherent state $\ket{\alpha}$. Thus, the initial state is given
as
\begin{equation}
  \label{eq:21}
  \op\rho(0)
  =
  \left(\begin{array}[c]{cc}
      c & 0 \\
      0 & 1-c 
    \end{array}\right)
  \otimes
  \keto{\alpha}\brao{\alpha}
\end{equation}
where the spin's state is given in its energy eigenbasis. The
self-generated process imposed on the spin via coupling to the
oscillator might thus be called ``adiabatic''; however, due to quantum
mechanical interactions the local entropy (purity) will, in general,
not be constant, see below.

\subsection{$z$-interaction ($z$-SOM)}
\label{sec:model:zint}

Representing the Hamiltonian (\ref{eq:Hsom}) in the eigenbasis of the
spin, one finds that it has a block-diagonal structure:
\begin{eqnarray}
  \Hopm{}
  & = &
  \left(
    \begin{array}[c]{cc}
      \Hop{o} - \lambda \op x - \frac{\om{S}}{2} & \\
      & \Hop{o} + \lambda \op x + \frac{\om{S}}{2} 
    \end{array}
  \right) \nonumber\\
  \label{eq:22}
  & =: &
  \left(
    \begin{array}[c]{cc}
      \Hop{-} & \\
      & \Hop{+} 
    \end{array}
  \right)
\end{eqnarray}
The same obviously holds for the time-evolution operator and -- by the
block-diagonal structure of the initial state -- also for the
propagated state $\op\rho(t)$ of the total system:
\begin{equation}
  \label{eq:23}
  \op\rho(t)
  =
  \left(
    \begin{array}[c]{cc}
      c \keto{\alpha_-(t)}\brao{\alpha_-(t)}  & \\
      & (1-c) \keto{\alpha_+(t)}\brao{\alpha_+(t)}
    \end{array}
  \right)
\end{equation}
Here we have used the definitions $\keto{\alpha_\pm(t)}:=\op
U_\pm(t,0)\keto{\alpha}$ and $\op U_\pm(t_1,t_0) =
\exp{[-i\Hop{\pm}(t_1-t_0)]}$. Note that the dynamics of the system
are periodic, because both Hamilton operators $\Hop{\pm}$ describe
(displaced) harmonic oscillators with the same frequency
$\om{o}$. Thus, we have $\op U(t_1+2\pi m \om{o}^{-1},t_0+2\pi n
\om{o}^{-1}) = \op U(t_1,t_0)$ for integer numbers $n,m$.

\begin{figure}
  \includegraphics[width=8cm]{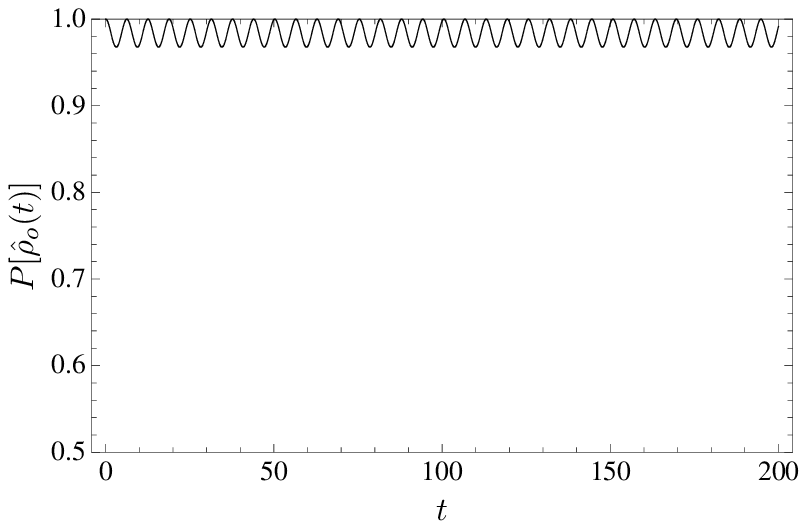}
  \caption{ \label{fig:1} Purity dynamics of the oscillator in the
    $z$-SOM for the special parameters $\lambda=0.1, c=0.7, \alpha=0,
    m=\om{s}=\om{o}=1$.}
\end{figure}

Because of the simple structure of the time-evolution of the system,
the purity of the oscillator can be computed analytically and turns
out to be given by
\begin{equation}
  \label{eq:24}
  P[\op\rho_o(t)]
  =
  c^2 + (1-c)^2 + 2 c(1-c)|\braketo{\alpha_-(t)}{\alpha_+(t)}|^2
\end{equation}
with the time-dependent part
\begin{equation}
  \label{eq:25}
  |\braketo{\alpha_-(t)}{\alpha_+(t)}|^2  
  =
  \exp\left[-8\frac{\lambda^2}{m\om{o}^3}\sin^2\left(\frac12\om{o}t\right)\right]
\end{equation}
(for the derivation, see Appendix~\ref{app:posc}; $m$ is the
oscillator mass). For pure initial spin states ($c=0,1$) we have
$P[\rop{o}(t)]=1$. An example for $P[\rop{o}(t)]$ for a mixed initial
spin state is given in Fig.~\ref{fig:1}.

It is easy to see from Eq.~(\ref{eq:24},\ref{eq:25}) that the minimum
purity with respect to $t$ and $c$ is
\begin{equation}
  \label{eq:26}
  P_o^\mathrm{min}
  =
  \frac12\left[1+\exp\left(-8\xi\right)\right]
  ,
\end{equation}
where
\begin{equation}
  \label{eq:68}
  \xi = \frac{\lambda^2}{m\om{o}^3}
  .
\end{equation}
Therefore, one has to choose $\xi\rightarrow 0$ and thus
\begin{equation}
  \label{eq:27}
  1-P_o^\mathrm{min}\ll 1
\end{equation}
in order to apply the FA. 

We can distinguish two different ways to enforce the limit $\xi
\rightarrow 0$:
\begin{gather}
  \label{eq:71}
  m\rightarrow\infty, \om{o}=\mathrm{const.}, \lambda=\mathrm{const.} \\
  \label{eq:73}
  \om{o}\rightarrow\infty, m=\mathrm{const.}, \lambda^2/\om{o}=\mathrm{const.}
\end{gather}
Their relevance will become clear in Sec.~\ref{sec:z-work}. If one
accepts the resulting finite $P_o^\mathrm{min}$ for some finite $\xi$,
the local coherence-time may be called infinite.

\subsection{$xz$-interaction ($xz$-SOM)}
\label{sec:model:xzint}

We discuss now the more complicated case of an interaction of form
Eq.~(\ref{eq:20}). This interaction is motivated by the following
considerations. First, the above case is very special in that the
minimum purity reached can be controlled completely by the system
parameters. The more general $xz$-interaction case will show that this
property is dependent on the interaction. Second, the chosen
interaction allows us to study the effect of imperfect control over
the exact form of the interaction as it might be the case for a more
realistic experimental situation. Finally, the $xz$-interaction
exhibits a remarkable diversity of dynamics which serves to illustrate
the pros and cons of the proposed work reservoir quality measures as
well as the possibility to realize quantum work sources within the
given model.

First, we show that a Hamiltonian
\begin{equation}
  \label{eq:HxzSOM}
  \Hopm{}
  =
  \frac{\om{s}}{2} \op \sigma_z + \lambda \op s \op x + \om{o}\left(\op a^\dagger \op a + \frac12\right)
\end{equation}
with an arbitrary operator $\op s$ acting on the spin's Hilbert space
is equivalent to the $xz$-SOM. This is seen from the expansion of $\op
s$ in the operator basis $\{\op
Q_k|k=0,1,2,3\}=\frac{1}{\sqrt{2}}\{\op 1, \op\sigma_x, \op\sigma_y,
\op\sigma_z\}$, which reads
\begin{equation}
  \label{eq:28}
  \op s = \sum\limits_{k=0}^3 \tr(\op Q_k^\+ \op s) \op Q_k
\end{equation}
(for details refer to \cite{GMVAW98}, pp. 34--49). The $\op 1$ term is
local to the oscillator and can be absorbed in $\Hop{o}$, while the
$\op\sigma_k$ terms can always be transformed to the form $\op s =
\lambda(\op\sigma_z+\kappa\op\sigma_x)$ by the local transformation
$\exp(-i \phi \op \sigma_z)$ with appropriately chosen
real~$\phi$. This corresponds to a rotation around the $z$-axis of the
spin.  The $xz$-SOM discussed here is therefore representative for the
whole class of Hamiltonians of the form Eq.~(\ref{eq:HxzSOM}).

\subsection{Dynamics}
\label{sec:model:xzdyns}

Now, we want to look into the behavior of the system for $|\kappa|
\lesssim 1$. To get insight into the dynamics, we invoke a rotating
wave approximation (RWA). For that purpose, we first write the
$xz$-SOM interaction Hamiltonian $\Hop{xz}$ in the
interaction picture
\begin{multline}
  \label{eq:29}
  \op{\tilde{H}}_{xz} \propto \op\sigma_z\op a
  \exp\left(i\frac{\Omega-\Delta}{2} t\right) + \op\sigma_z\op a^\+
  \exp\left(-i\frac{\Omega-\Delta}{2} t\right) \\+
  \lambda\op\sigma_+\op a \exp(-i\Delta t) + \lambda\op\sigma_-\op a
    \exp(i\Omega t) \\
    +\lambda\op\sigma_+\op a^\+ \exp(-i\Omega t)
    +\lambda\op\sigma_-\op a^\+ \exp(i\Delta t)],
\end{multline}
where we have defined $\Omega:=\om{s}+\om{o}$ and
$\Delta:=\om{s}-\om{o}$. By restricting ourselves to the resonant case
$\Delta=0$ and omitting all terms rotating with frequencies $\Omega$
and $\Omega/2$, the $xz$-SOM Hamiltonian in RWA turns out to be
\begin{equation}
  \label{eq:30}
  \Hop{xz}^\mathrm{RWA}
  = 
  \frac{\omega}{2} \op\sigma_z + g ( \op\sigma_+\op a + \op\sigma_-\op a^\+ ) + \omega\left( \op a^\+ \op a + \frac12 \right)
\end{equation}
in the Schrödinger picture. This is just the Hamiltonian $\Hopm{JC}$
of the Jaynes-Cummings model (JCM) \cite{JC63,SK93} with 
\begin{equation}
  \label{eq:31}
  g = \frac{\lambda \kappa}{\sqrt{2 m \omega}}
\end{equation}
and $\omega=\om{s}=\om{o}$. According to \cite{Shi65,Lar07}, the RWA
is accurate as long as $g/\Omega\ll 1$. This condition is met for all
relevant cases, since we consider in the following a situation where
the parameters $m,\om{}$, and $\lambda$ have been chosen such that for
$\kappa=0$ ($z$-SOM) the FA holds for all times.

Now let us turn to the interpretation of the result. First, we note
that by performing the RWA, in particular, the $\op\sigma_z$ term of
the interaction Hamiltonian is removed. Therefore, the $xz$-SOM in RWA
captures the effect of the $\op\sigma_x$ interaction alone and, in
turn, this means that the main dynamics are governed by the
$\op\sigma_x$ part of the interaction alone.

Second, the dynamics of the JCM (and therefore of the $xz$-SOM in RWA)
scale in time with $g^{-1}$. This is most clear from the exact
time-evolution operator for the JCM (\cite{MOS97}, p.~205) in the case
of exact resonance
\begin{equation}
  \label{eq:32}
  \op U_\mathrm{JC}(t)
  = 
  \left(
    \begin{array}[c]{cc}
      \cos( g t \op B ) & - i \op a^\+ \sin(gt\op A)\op A^{-1} \\ 
      -i \sin(gt\op A)\op A^{-1}\op a & \cos(g t \op A) 
    \end{array}
  \right)
\end{equation}
with $\op A=\sqrt{\op a^\+ \op a + 1}$ and $\op B = \sqrt{\op a^\+ \op
  a}$.

\begin{figure}
  \includegraphics[width=8cm]{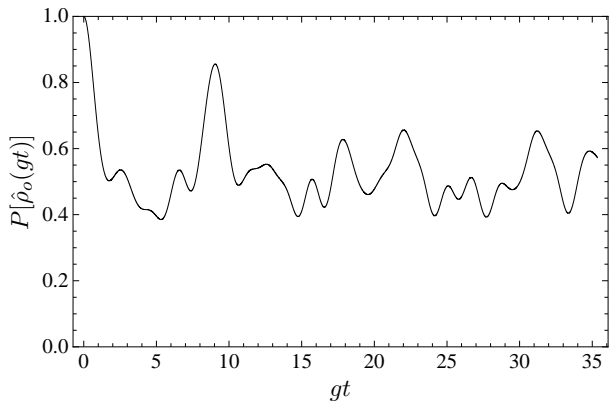}
  \caption{\label{fig:2} Purity of the oscillator for numerically
    exact dynamics of the $xz$-SOM in RWA for arbitrary $g$. The
    parameters are: $\lambda=0.01, \alpha=\om{o}=\om{s}=1$, $c=0.7$.}
\end{figure}

\begin{figure}
  \begin{tabular}{c}
    \includegraphics[width=8cm]{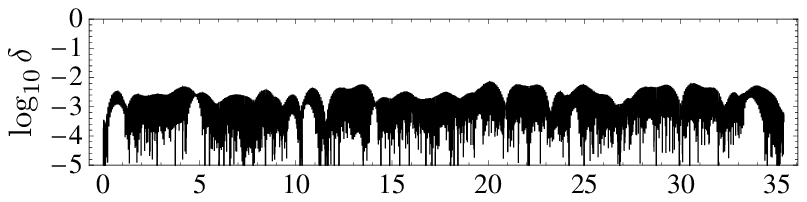}\\
    \includegraphics[width=8cm]{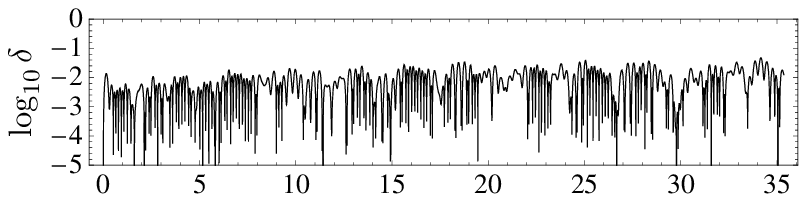}\\
    \includegraphics[width=8cm]{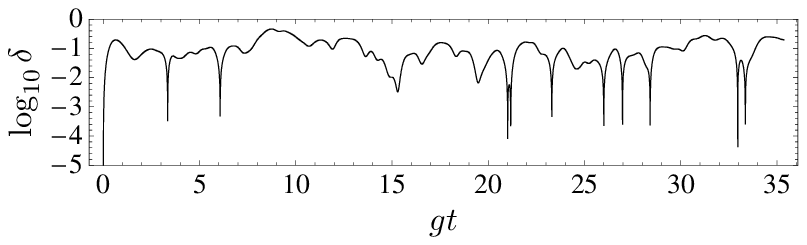}\\  
  \end{tabular}
  \caption{\label{fig:3} Common logarithm of the absolute deviation
    $\delta$ of the (numerically exact) purity dynamics with and
    without RWA for (top to bottom) $g=-0.01,-0.1,-1$ corresponding to
    $\kappa=\sqrt{2},10\sqrt{2},100\sqrt{2}$ according to
    Eq.~(\ref{eq:31}) with the parameters
    $\lambda=-0.01,m=1,\alpha=\om{o}=\om{s}=1$, $c=0.7$.}
\end{figure}

The numerical results of the dynamics of the purity of the oscillator
are given in Fig.~\ref{fig:2} for the case of a coherent initial state
with one photon in the cavity on average ($\alpha=1$). The deviations
of the numerical results for the $xz$-SOM with and without RWA for
three different orders of magnitude of $g$ are given in
Fig.~\ref{fig:3}.

From Fig.~\ref{fig:3} one sees that the RWA yields good results (less
then 10\% relative deviation) up to $\kappa\approx10$. This shows
again that the RWA gives an accurate description of the $xz$-SOM
dynamics in agreement with the expectation given above.

It is obvious from Figs.~\ref{fig:1},\ref{fig:2} that the purity
behavior of the $xz$-SOM is fundamentally different from the
$z$-case. The decrease of the purity due to the additional
$\op\sigma_x$ interaction term is several orders of magnitude stronger
than what is expected from the $\op\sigma_z$ term alone and due to the
approximate scaling behavior of the $xz$-SOM, the minimum does not
depend on $\kappa$, as long as $\kappa$ is not zero. We conclude from
that, that in the presence of an arbitrarily small but non-vanishing
$\op\sigma_x$ term the FA and with it the work reservoir quality of
the oscillator will break down in finite time. Reduction of $\kappa$
can only delay the breakdown and, thus, if $\lambda,\kappa\neq0$ no
choice of the other model parameters can prevent the breakdown. This
result is in agreement with the finding in \cite{gemmer0108} for the
that the coherence time for the JCM depends on the interaction
strength such that weaker interaction leads to longer coherence time.

In the above sense, the work reservoir functionality in the given
quantum scenario is quite sensitive to the quality of control of the
interaction between spin and oscillator.

\section{Work source quality in the $z$-SOM}
\label{sec:z-work}

Considering the first case of $z$-SOM, let us assume that one has
chosen the parameters of the system such that Eq.~(\ref{eq:27}) is
fulfilled. We can then apply the FA not only to describe the dynamics
of the system up to any desired accuracy but moreover, we get a new
level of description combined with new physical insight in the
properties and characteristics of the system. This will be outlined
below.

Applying the FA to the SOM, we find according to
Eqs.~(\ref{eq:genFAmixed}) and (\ref{eq:genFApure}) the following
effective coupled equations
\begin{eqnarray}
  \label{eq:33}
  i \dot{\op{\rho_s}} & = & \left[ \left(\frac{\om{s}}2 + \lambda \expv{\op x}(t)\right) \op\sigma_z, \op\rho_s \right] \\
  \label{eq:34}
  i \keto{\dot{\Psi}(t)} & = & \left( \Hop{o} + \lambda (1-2 c)\op x \right) \keto{\Psi(t)}
  ,
\end{eqnarray}
where we have defined $\expv{\op x}(t):=\brao{\Psi(t)}\op
x\keto{\Psi(t)}$. Hence, the spin is driven by the oscillator
displacement which acts like an additional time-dependent magnetic
field, modulating the spin's Zeeman splitting. On the other hand, the
spin dynamics lead to a constant displacement of the oscillator
potential. In this sense there is asymmetry between the two
subsystems: The effective Hamiltonian for the oscillator is modified
but not time-dependent.

Clearly, this result is in agreement with the result which one would
obtain from applying the integral measure based on $\mathcal{W}$ and
$\mathcal{Q}$: The latter is zero, since according to
Sec.~\ref{sec:model:zint} and Eq.~(\ref{eq:33}), the spin's state does
not change and the oscillator only exerts classical driving on the
spin. The local effective energy change of the spin resulting from the
driving is 100\% work. Hence, the oscillator acts as an ideal work
source at least in the limits discussed in Sec.~\ref{sec:model:zint},
Eqs.~(\ref{eq:71},\ref{eq:73}).

However, it has to be noted that the peak-to-peak amplitude $\Domseff
= \lambda\left(\expv{\op x}_\mathrm{max}-\expv{\op
    x}_\mathrm{min}\right)$ of the effective spin splitting is
dependent on the system parameters (see App.~\ref{app:stroke}):
\begin{equation}
  \label{eq:64}
  \Domseff = 2 \lambda \sqrt{\frac{2}{m\om{o}}} |\alpha + \gamma|,
\end{equation}
where $\gamma = \sqrt{\lambda^2/(2m\om{o}^3)}(1-2c) =
\sqrt{\xi/2}(1-2c)$ [cf. Eq.~(\ref{eq:68})]. For the first limit
proposed in Eq.~(\ref{eq:71}), we therefore find that both
$\gamma\rightarrow 0$ and $\Domseff\rightarrow 0$, if all parameters
besides $m$ are kept constant. Thus, the work effect induced by the
oscillator diminishes more and more for increasingly better fulfilled
FA. This can be avoided, though, by additionally imposing
$\alpha\rightarrow\infty$, such that $|\alpha|^2/m$ remains constant,
which then defines the splitting's amplitude. This is a classical
limit in that the mass and average excitation number of the oscillator
go to infinity.

There is also a true quantum limit, though, which is found to be
realized exploiting the second limit given in Eq.~(\ref{eq:73}). Here,
by letting $\om{o}\rightarrow\infty$, we enforce that $\xi\rightarrow
0$ so that the factorization approximation becomes exact. However, by
requiring $\lambda^2/\om{o} = \mathrm{const.}$, the prefactor of
$\Domseff$ in Eq.~(\ref{eq:64}), $2\sqrt{2\lambda^2/(m\om{o})}$,
becomes constant.  Thus, although $\gamma\rightarrow 0$, $\Domseff$
retains a finite value
\begin{equation}
  \label{eq:74}
  \Domseff \rightarrow 2 \lambda \sqrt{\frac{2}{m\om{o}}} |\alpha| = \mathrm{const.}
\end{equation}
for arbitrary (small but finite) $m$ and $\alpha$ in the limit of
exact FA.

By the preceding reasoning, we conclude that the oscillator is,
indeed, a work reservoir for the spin, periodically changing the spin
splitting and therefore transferring work to/from it. What is special
about that finding is the fact that a true quantum system (the
oscillator in the quantum limit of the $z$-SOM) can be set up as an
ideal work source and, thus, the work concept is not tied to classical
devices. Moreover, as long as we fulfill Eq.~(\ref{eq:27}) the
oscillator behaves as a work reservoir for any time-period.

\section{Work source quality in the $xz$-SOM}
\label{sec:xz-work}

In this section, we present and discuss our results for the work
reservoir behavior in the more general $xz$-SOM presented in
Sec.~\ref{sec:model:xzint} with focus on the suitability of the work
source quality measures proposed in Sec.~\ref{sec:q-meas}.

The numerical results used herein have been produced with the
\emph{Mathematica} package using the following techniques: We have
computed the time evolution of the system by direct diagonalization of
the Hamilton operator (with a cut-off chosen such that only states
with occupation probability higher than $10^{-6}$ are
included). Integration of quantities -- where necessary -- has been
performed using the rectangle rule and the error of integration has
been controlled by crosschecking with results for the trapezoidal rule
and/or for smaller time steps.

\subsection{Purity based approach}
\label{sec:xz-work:purity}

We need now to define a lower bound for the purity of the
oscillator. The $\op\sigma_z$ coupling alone already leading to some
limited purity loss in the oscillator can be considered as sort of a
``natural'' purity drop, which has to be accepted for any system that
interacts at all and that is present even if the FA is good and the
work source quality high. 

The work source functionality is considered to fail when the purity
decrease of the RWA dynamics (caused by the $\op\sigma_x$ interaction
term) reaches the maximum purity drop $P_o^\mathrm{min}$ of the
$\op\sigma_z$ dynamics alone found for the given system parameters
($\lambda, \om{s}, \om{0}, m$). This allows us to define a breakdown
time $t^*$ by $P[\rop{o}(t^*)]=P_o^\mathrm{min}$. 

The close connection of the $xz$-SOM dynamics to the JCM dynamics
seems to suggest an analytical approach based on the standard
approximations made to solve the JCM (see, e.g., \cite{BD00}, Ch.~6):
approximation of the occupation probability of the coherent state with
a Gaussian and linearization of the spectrum of the JCM around its
peak. With those approximations a fairly accurate description of the
JCM's typical collapse and revival behavior of the spin polarization
for high initial photon numbers ($|\alpha|\gg 1$) is possible. After
the initial collapse of polarization, the spin reaches its minimum
purity \cite{SK93} and the oscillator purity will as well have dropped
significantly. 

Unfortunately, even in the high photon number limit the accuracy of
this approach in the relevant time interval up to this point of the
evolution is insufficient: Typical values of the purity drop due to
the $\op\sigma_z$ interaction are of the order of $10^{-2}$, while the
error of the mentioned approximations is of around the same order
during the collapse. This renders the application of those
approximations futile and since a full analytical analysis is much too
involved, we will only exemplify some results based on numerics.

To this end, we choose the following parameters for the $xz$-SOM:
$\om{o}=\om{s}=1$ (resonant case), $m_o=1$, $\lambda=\kappa=0.1$. In
the following, we consider the results of two special cases:
\begin{enumerate}
\item[(a)] $\alpha=0,\, c=0.5$
\item[(b)] $\alpha=2,\, c=1$
\end{enumerate}
These two examples are drawn from a set of results for initial states
with parameters $\alpha\in[0,4]$ and $c\in[0.5,1.0]$ and have been
chosen for they represent in some sense extremal cases, that will be
seen to illustrate the features of the different work source quality
measures. A short overview about the more general behavior of the
$xz$-SOM is given in Sec.~\ref{sec:xz-work:overall}.

\begin{figure}
  \includegraphics[width=\linewidth]{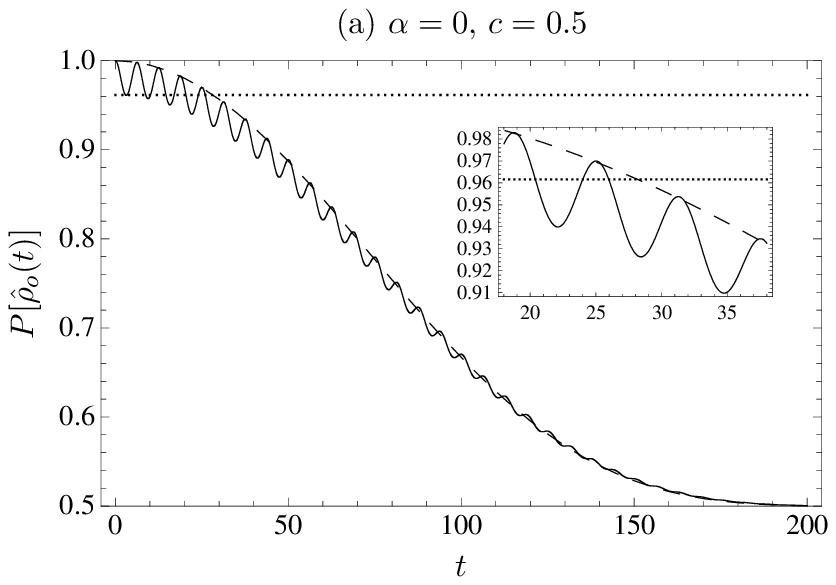}
  \phantom{x}
  \includegraphics[width=\linewidth]{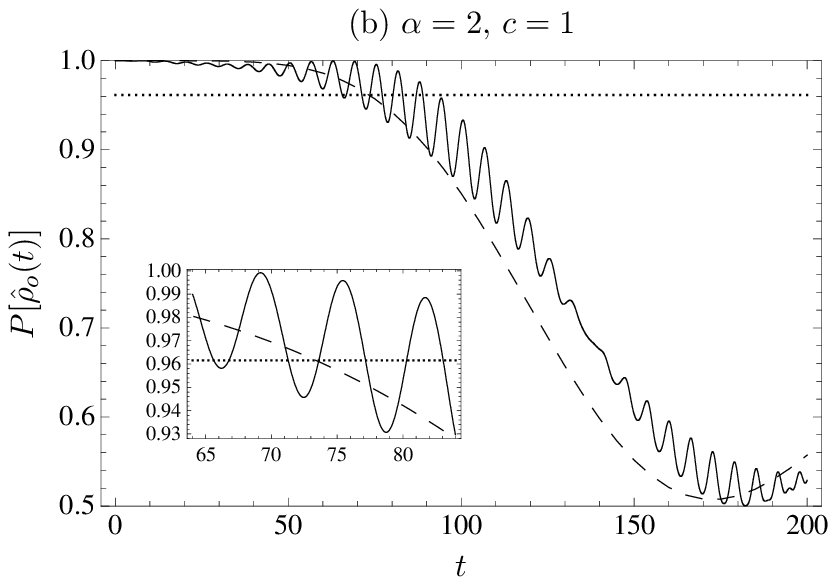}
  \caption{\label{fig:4} Oscillator purity behavior of two special
    cases of $xz$-SOM and comparison with minimum purity of $z$-SOM
    for the given system parameters: Numerical exact result (solid
    line), numerical result with RWA (dashed), minimum $z$-SOM purity
    (dotted), which -- according to Eq.~(\ref{eq:26}) -- is
    $[1+\exp(-2/25)]/2\approx 0.962$. The insets show the crossings of
    RWA-purity with minimum $z$-SOM purity.}
\end{figure}

The purity behavior of the examples is shown in Fig.~\ref{fig:4}. The
time after which the FA is estimated to fail is roughly
$t^*_{(a)}\approx28$ and $t^*_{(b)}\approx73$. Although this means
that the second case is expected to exhibit work reservoir
functionality about three times longer than the first case, one would
conclude from the curves that for both cases, the oscillator's work
source functionality degrades quickly after the initial high purity
phase and is virtually absent at least for $t>100$.

\subsection{Work/heat based approach}
\label{sec:xz-work:wq}

\begin{figure}
  \includegraphics[width=\linewidth]{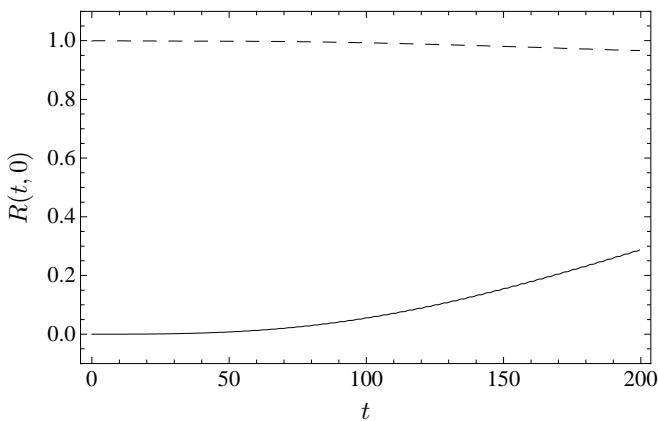}
  \caption{\label{fig:5} Results of the integral work reservoir
    quality measure $R(t,0)$ for the examples as of Fig.~\ref{fig:4}:
    (a)~$\alpha=0,c=0.5$ (solid line), (b)~$\alpha=2,c=1$ (dashed)}
\end{figure}

However, taking a look at the result for the integral quality measure
$R$ shown in Fig.~\ref{fig:5} one comes to a completely different
conclusion: In case (a) the oscillator starts as a perfect heat source
rather than a perfect work source and only in the course of time a
work source effect arises, whereas in case (b) the oscillator is
recognized as a nearly ideal work source during the whole interval. It
is astonishing to see that the purity based measure gives such a
different picture since the reasoning based on the FA is valid: During
the initial phase, $\Leff{s}$ is close to 0.

\begin{figure}
  \includegraphics[width=\linewidth]{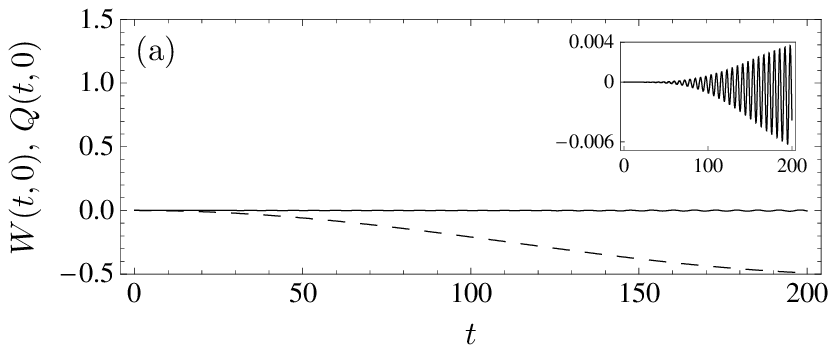}
  \includegraphics[width=\linewidth]{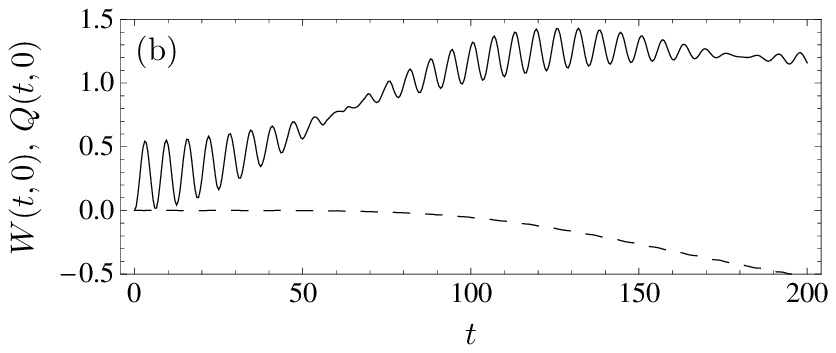}
  \caption{\label{fig:6} Integrated work $W(t,0)$ (solid line) and
    heat $Q(t,0)$ (dashed) for the two chosen examples, as of
    Fig.~\ref{fig:5}. The inset of case (a) shows $W(t,0)$ alone.}
\end{figure}

The reason for this seemingly contradictory characterization becomes
evident when examining the results for the integrated work and heat,
$W(t)$ and $Q(t)$ (see Fig.~\ref{fig:6}). In both presented cases, the
total heat flow in the beginning of the dynamics is small as expected
from the FA argument. Also, the heat becomes significant not before
$t^*_{(a)}$ and $t^*_{(b)}$, respectively, which again demonstrates
the strong connection between heat and purity.

However, the work exhibits a completely different behavior for the two
cases: In the first case, the work remains almost constant at zero
until oscillations set in at around $t\approx100$ (see the inset of
Fig.~\ref{fig:6}(a)). Those oscillations lead to the slow increase in
work source quality in the second half of the considered time
interval. Although the oscillations have only small amplitude, their
work source effect becomes significant due to their frequency, which
is high when compared to the time scale of heat dynamics.

In case (b), $W(t,0)$ shows strong oscillations of an amplitude orders
of magnitude larger than in case (a) from the very beginning of the
dynamics. Thus, the reason for the contradiction to the purity measure
result is traced back to the problem already touched on in
Sec.~\ref{sec:xz-work:purity}: Although the purity can be used as a
measure for the size of incoherent part of the effective dynamics of
the spin, $\Leff{s}$, which is associated with the heat flow, it is
completely insensitive to the size and effects of the coherent par
$\Heff{s}$ and thus the actual work source effect.

From this result, we can draw the conclusion that due to the fact that
the purity measure is only linked to $\Leff{s}$ alone, it can only be
used as a necessary condition for work source functionality. In order
to get the full picture, a more detailed analysis of the work and heat
fluxes via the measure $R$ is necessary.

\subsection{Work source quality overview}
\label{sec:xz-work:overall}

Computing $R$ for a low photon number parameter window
($\alpha \in [0,4]$) and initial spin temperatures ranging from 0 to
$\infty$ ($c \in [0.5,1]$), we find the following trends: 

For $\alpha=0$, the overall work source quality of the considered
interval $t \in [0,200]$ is generally significantly lower than for the
corresponding (with respect to $c$) cases for $\alpha>0$, and
$R(200,0)$ ranges from 0.8 to 0.2 with decreasing $c$. For $\alpha>0$,
$R(200,0)$ takes on values around 0.9, with a slow increase for higher
$\alpha$ and $c$. The first increase can be related to the higher
excitation of the oscillator and the resulting bigger amplitude of the
position expectation. The second trend has to do with a special
property of the initial states $\ket{0}\ket{\alpha}$ and
$\ket{1}\ket{\alpha}$ of which the initial state of the $xz$-SOM is a
statistical mixture.

In order to explain the trend for increasing $c$, we invoke the first
order perturbation theory for the extremal initial states
$\ket{0}\ket{\alpha}$ and $\ket{1}\ket{\alpha}$ which is applicable to
the beginning of the dynamics, as long as $gt^*\ll 1$ holds with
$g=10^{-2}/\sqrt{2}$. The calculation is carried out in
App.~\ref{app:1tdpt}. Here, we only make use of the result
\begin{align}
  \label{eq:35}
  \op U^I(t)\ket{0}\ket{\alpha}  =& \left( \ket{0} - i \alpha g t \ket{1} \right) \ket{\alpha} + \mathcal{O}(g^2)\\
  \begin{split}
    \op U^I(t)\ket{1}\ket{\alpha}  = &\left( \ket{1} - i \frac{\alpha^*}{2} g t \ket{0} \right) \ket{\alpha} \\ 
    &- igt\ket{0}\frac{\partial}{\partial \alpha^*}\ket{\alpha}+ \mathcal{O}(g^2)
  \end{split}
\end{align}
and $\alpha^*$ denotes the complex conjugate of $\alpha$. From this
form of the state $\partial_{\alpha^*} \ket{\alpha}$, we easily see
that in first order $\op U^I(t)\ket{0}\ket{\alpha}$ factorizes
contrary to $\op U^I(t)\ket{1}\ket{\alpha}$. The purity behavior of
the initial state
$\rho(0)=[c\ketbra{0}{0}+(1-c)\ketbra{1}{1}]\otimes\ketbra{\alpha}{\alpha}$
continuously changes from the $\ket{0}\ket{\alpha}$ case to the
$\ket{1}\ket{\alpha}$ case. As the $\ket{1}\ket{\alpha}$ state becomes
more and more mixed into the initial state with decreasing $c$, a
significant purity drop happens at earlier times of the evolution. The
same is true for the heat flow, which is tied to the purity drop. With
this increased heat flow at the early stage of the evolution,
$\mathcal{Q}(200,0)$ reaches higher values for decreasing $c$.

Moreover, the size of work flux for the same change of effective
splitting of the spin decreases with decreasing $c$ until it reaches 0
for $c=0.5$. Thus, in the beginning and as long as the spin's state is
close to its initial occupation, the work source effect of the
oscillator is reduced or suppressed additionally. Clearly, this
reduces the work source quality and explains the trend seen in the
numerical results.


\section{Summary}
\label{sec:sum}

Work and heat are related to (thermodynamic) processes, which seem to
require external control. In this paper, we have argued that work
functionality may show up in closed bipartite quantum systems, even
down to the nanoscale. We have shown under what conditions the
respective subsystem dynamics may be described via time-dependent
effective Hamiltonians and in this sense act as classical driver for
each other.

We have then brought forward the argument that energy exchanged under
such conditions has to be considered as work from the viewpoint of
thermodynamics. We have applied these results to a simple model
confirming that a system as small and simple as a single harmonic
oscillator coupled to a spin can act as a work reservoir for the
latter. In addition, we have introduced purity based and work/heat
based work source quality measures and discussed their usefulness. We
have demonstrated that due to the lack of sensitivity to the effects
of $\Heff{1}$, the purity based measure is only a necessary condition
for work source functionality. In general, the implementation of a
full thermodynamic process within a closed quantum system will require
driving as well as thermalizing embeddings.

We would like to thank A. Allahverdyan, A. Kettler, K. Rambach,
F. Rempp, J. Teifel, P. Vidal, G. Waldherr, H. Weimer and M. Youssef
for fruitful discussions. Financial support of the Studienstiftung des
Deutschen Volkes is gratefully acknowledged.

\appendix
\section{Generalized factorization approximation}
\label{app:gfa}

In its original form the FA is formulated for a bipartite system
(cf. \cite{gemmer0108}, Eq.~(\ref{eq:3},\ref{eq:4})). Let us now consider a
tripartite system defined by the Hamiltonian
\begin{equation}
  \label{eq:36}
  \op H = \op H_0 + \op H_1 + \op H_{12} + \op H_2 \, .
\end{equation}
Let us assume that system 0 has interacted with system 1 in the past
but is now decoupled from system 1. System 2, however, is supposed to
have been uncoupled in the past and is now being coupled to system 1
alone. Finally, we assume that the combined system 01 and system 2 are
now in a pure state. We are then left with an initial state for the
whole system of the form
\begin{equation}
  \label{eq:37}
  \op \rho(0) = \op \rho_{01}(0) \otimes \op \rho_2(0)
\end{equation}
with $\op \rho_{01}(0) = \ketbra{\Psi_{01}(0)}{\Psi_{01}(0)}$ and $\op
\rho_{2}(0) = \ketbra{\Psi_{2}(0)}{\Psi_{2}(0)}$.

We now consider the dynamics of this system with respect to the given
initial state. In the case that $\tr[\op\rho_2^2(t)]\approx 1$ holds,
the FA is applicable to the whole system yielding the two coupled
equations
\begin{align}
  i \hbar \ket{\dot\Psi_{01}(t)} =& \left(\Hop{0}+\Hop{1}+\right. \nonumber \\
  \label{eq:38}
  &\left. + \bra{\Psi_2(t)}\Hop{12}\ket{\Psi_2(t)} \right) \ket{\Psi_{01}(t)} \\
  \label{eq:39}
  i \hbar \ket{\dot\Psi_2(t)} =&\left(\Hop{2}+\bra{\Psi_{01}(t)}\Hop{12}\ket{\Psi_{01}(t)}\right)\ket{\Psi_2(t)} \, .
\end{align}
By restating Eq.~(\ref{eq:38}) in the form
\begin{equation}
  \label{eq:40}
  i \hbar \dot{\op{\rho}}_{01}(t) = \left[\Hop{0}+\Hop{1}+\bra{\Psi_2(t)}\Hop{12}\ket{\Psi_2(t)},\op\rho_{01}(t)\right]
\end{equation}
and taking the trace of the Hilbert space of the ancillary system 0,
we arrive at the result
\begin{equation}
  \label{eq:41}
  i \hbar \dot{\op{\rho}}_1(t) = \left[\Hop{1}+\bra{\Psi_2(t)}\Hop{12}\ket{\Psi_2(t)},\op\rho_1(t)\right] \, .
\end{equation}
To get this result, we have made use of the two partial trace
relations
\begin{eqnarray}
  \label{eq:42}
  \tr_0 \left[\op A\otimes\op1_1, \op B\right] &=& 0 \\
  \label{eq:43}
  \tr_0 \left[\op1_0\otimes\op A, \op B\right] &=& \left[\op A, \tr_0 \op B\right]
\end{eqnarray}
Note that in contrast to the case of the FA for a bipartite system,
the criterion for the applicability of the FA is the purity dynamics
of system 2 alone.

\section{Purity of the oscillator}
\label{app:posc}

The purity dynamics of the oscillator in the case of the pure
$\op\sigma_z\op x$ interaction can be derived from the solution of the
Liouville-von Neumann equation given in Eq.~(\ref{eq:23}),
\begin{equation*}
  \op\rho(t)
  =
  \left(
    \begin{array}[c]{cc}
      c \keto{\alpha_-(t)}\brao{\alpha_-(t)}  & \\
      & (1-c) \keto{\alpha_+(t)}\brao{\alpha_+(t)}
    \end{array}
  \right) .
\end{equation*}
Thus, the oscillator reduced state is
\begin{equation}
  \label{eq:44}
  \op\rho_o(t) =  c \keto{\alpha_-(t)}\brao{\alpha_-(t)} + (1-c) \keto{\alpha_+(t)}\brao{\alpha_+(t)}
\end{equation}
and taking the square and the trace of this expression, we end up with
the result for the purity given in Eq.~(\ref{eq:24}),
\begin{equation*}
  P[\op\rho_o(t)]
  =
  c^2 + (1-c)^2 + 2 c(1-c)|\braketo{\alpha_-(t)}{\alpha_+(t)}|^2 .
\end{equation*}
For the time-dependent term we find
\begin{gather}
  \label{eq:45}
  | \braketo{\alpha_-(t)}{\alpha_+(t)} |^2 = | \brao{\alpha}\op U^\dagger_-(t,0)\op U_+(t,0)\keto{\alpha} |^2 \\
   = | \brao{\alpha}\exp(i \op H_- t)\exp(-i\op H_+ t)\keto{\alpha} |^2 
\end{gather}
with $\op H_\pm = \op H_o \pm \lambda \op x \pm
\frac{\om{s}}{2}$. Making use of the displacement operator $\op
D(\alpha) = \exp(\alpha \op a^\dagger - \alpha^* \op a )$ and its properties
\begin{eqnarray}
  \label{eq:46}
  \op D(-\alpha) \op x \op D(\alpha) &=& \op x + \sqrt{\frac{2}{m\om{o}}}\Re(\alpha) \\
  \label{eq:47}
  \op D(-\alpha) \op p \op D(\alpha) &=& \op p + \sqrt{2m\om{o}}\Im(\alpha),
\end{eqnarray}
we can express $\op H_\pm$ as
\begin{equation}
  \label{eq:48}
  \op H_\pm = \op D(-\beta_\pm) \op H_o \op D(\beta_\pm) + C
\end{equation}
and therefore have
\begin{equation}
  \label{eq:49}
  \op U_\pm(t,0) = \ee^{-i C t} \op D(-\beta_\pm) \exp(-i \op H_o t) \op D(\beta_\pm)
\end{equation}
with $\beta_\pm = \mp\lambda/\sqrt{2m\om{o}^3}$ up to constant factors
or a phase, respectively, which are irrelevant for the computation of
the modulus in Eq.~(\ref{eq:45}). With the help of the relations
\begin{eqnarray}
  \label{eq:50}
  \op D(\alpha) \op D(\beta) & = & \exp[i \Im(\alpha\beta^*)]\op D(\alpha+ \beta) \\
  \label{eq:51}
  \exp(-i\op H_o t) \ket{\alpha} & = & \exp(-i \om{o} t /2)\ket{\alpha \exp(-i\om{o}t)}
\end{eqnarray}
we arrive at 
\begin{equation}
  \label{eq:52}
  \ket{\alpha_\pm(t)} = \exp[i\phi_\pm(t)]\ket{(\alpha+\beta_\pm)\exp(-i\om{o}t)-\beta_\pm} .
\end{equation}
Finally making use of the relation $|\braket{\alpha}{\alpha'}|^2 =
\exp(-|\alpha-\alpha'|^2)$ yields the result
\begin{equation}
  \label{eq:53}
  |\braket{\alpha_-(t)}{\alpha_+(t)}|^2 = \exp\left[-8\frac{\lambda^2}{m\om{o}^3}\sin^2\left(\frac12 \om{o} t\right)\right] .
\end{equation}

\section{Amplitude of the spin's effective splitting in the $z$-SOM}
\label{app:stroke}

In the case of the $z$-SOM and applicable FA, the effective
Hamiltonians of the spin and oscillator are found to be
[Eqs.~(\ref{eq:33},\ref{eq:34})]
\begin{eqnarray}
  \label{eq:65}
  \Heff{s}(t) &=& \left(\frac{\om{s}}2 + \lambda \expv{\op x}(t)\right) \op\sigma_z \\
  \label{eq:66}
  \Heff{o} &=& \left( \Hop{o} + \lambda (1-2 c)\op x \right)
\end{eqnarray}
and the latter may be rewritten in the form
\begin{equation}
  \label{eq:67}
  \Heff{o} = \frac12\om{o}\left(\op{\tilde{X}}^2+\op{\tilde{P}}^2\right) + \mathrm{const.}
\end{equation}
with the dimensionless position and momentum operators
\begin{eqnarray}
  \label{eq:69}
  \op{\tilde{X}} &=& \op X + \sqrt{\frac{\lambda^2}{m\om{o}}}(1-2c) \\
  \label{eq:70}
  \op{\tilde{P}} &=& \op P,
\end{eqnarray}
where $\op X = \sqrt{m\om{o}}\op x$, $\op P = \op
p/\sqrt{m\om{o}}$. Making use of the properties of the displacement
operator $\op D(\alpha)$ in Eqs.~(\ref{eq:46},\ref{eq:47}), one finds
that 
\begin{equation}
  \label{eq:72}
  \Heff{o}=\op D(-\gamma)\Hop{o}\op D(\gamma)  
\end{equation}
with
\begin{equation}
  \label{eq:56}
  \gamma = \sqrt{\frac{\lambda^2}{2m\om{o}}}(1-2c).
\end{equation}
In order to compute the peak-to-peak amplitude of the effective spin
splitting
\begin{equation}
  \label{eq:57}
  \Domseff = \lambda\left(\expv{\op x}_\mathrm{max}-\expv{\op x}_\mathrm{min}\right)
\end{equation}
we need to evaluate
\begin{align}
  \label{eq:77}
  &\expv{\op x}(t) = \bra{\alpha(t)}\op x\ket{\alpha(t)} \\
  &= \bra{\alpha}\exp(i\Heff{o}t)\op x\exp(-i\Heff{o}t)\ket{\alpha} \notag \\
  &= \bra{\alpha}\op D(-\gamma)\exp(i\Hop{o}t)\op D(\gamma)\op x\op
  D(-\gamma)\exp(-i\Hop{o}t)\op D(\gamma)\ket{\alpha} \notag \\
  \label{eq:76}
  &= \bra{\alpha+\gamma}\exp(i\Hop{o}t)\op x\exp(-i\Hop{o}t)\ket{\alpha+\gamma} + \tilde\gamma,
\end{align}
where we have used Eqs.~(\ref{eq:46},\ref{eq:50},\ref{eq:72}) assuming
$\alpha \in \mathbb{R}$ and defining
$\tilde\gamma=\sqrt{2/(m\om{o})}\gamma$. Now, we can see that the
first term is just the time evolution of the expectation value of the
position of the original oscillator described by $\Hop{o}$ for a
coherent initial state $\ket{\alpha+\gamma}$. With the help of
Eq.~(\ref{eq:51}) it is straightforward to show that $\expv{\op
  X}_\mathrm{max}-\expv{\op X}_\mathrm{min}=2\sqrt{2}|\alpha+\gamma|$
and therefore
\begin{equation}
  \label{eq:78}
  \Domseff = 2 \lambda \sqrt{\frac{2}{m\om{o}}} |\alpha + \gamma| .
\end{equation}
Note that this result is only exact if the Hamiltonian governing the
oscillator's dynamics is $\Heff{o}$ and $\Leff{o}[\rop(t)]=0$, that is
if the FA is exact. Still, if the FA holds in good approximation,
Eq.~(\ref{eq:78}) is a good approximation as well.

\section{First order time-dependent perturbation theory for pure initial states of JCM}
\label{app:1tdpt}

It is convenient to apply the perturbation theory in the interaction
picture. All interaction picture quantities are denoted by a
superscript ``$I$''. The expansion of the time-evolution of the state
is given by
\begin{align}
  \label{eq:54}
  \ket{\Psi^I(t)} = \ket{\Psi(0)} + \op U_1^I(t)\ket{\Psi(0)} +
  \mathcal{O}(g^2)
\end{align}
and the first order contribution to the time-evolution operator $\op
U_1^I(t)$ is given by (see, e.g., \cite{Mes90}, p. 207ff)
\begin{align}
  \label{eq:55}
  \op U_1^I(t) = - i \int\limits_0^t \dd\tau \op V^I(\tau)
\end{align}
and 
\begin{align}
  &\op V^I(t) = \op U_0^\dagger(t) \op V \op U_0(t) \notag \\
  &= g\exp[i(\Hop{s}+\Hop{o})t](\op\sigma_+\op a +
  \op\sigma_-\op a^\dagger)\exp[-i(\Hop{s}+\Hop{o})t] \notag \\
  \label{eq:58}
  &= g(\op\sigma_+\op a + \op\sigma_-\op a^\dagger)
\end{align}
is the interaction operator in the interaction picture. According to
the RWA, only terms of the interaction are kept which are
time-independent in the interaction picture, thus the last
equality. From Eqs.~(\ref{eq:58}) and (\ref{eq:55}), it follows that
\begin{equation}
  \label{eq:59}
  \op U_1^I(t) = - i g t (\op\sigma_+\op a + \op\sigma_-\op a^\dagger) .
\end{equation}
The time evolution of a state in the JCM is therefore given in first
order perturbation by
\begin{equation}
  \label{eq:60}
  \ket{\Psi^I(t)} = [\op 1 - i g t (\op\sigma_+\op a + \op\sigma_-\op a^\dagger)] \ket{\Psi(0)} + \mathcal{O}(g^2) .
\end{equation}

Here, we consider $\ket{0}\ket{\alpha}$ and $\ket{1}\ket{\alpha}$ as
initial states. Together with
\begin{equation}
  \label{eq:61}
  \op a^\dagger \ket{\alpha} = \left(\frac{\partial}{\partial \alpha^*} + \frac{\alpha*}{2} \right) \ket{\alpha} 
\end{equation}
(see \cite{VW06}) we find for those states
\begin{align}
  \label{eq:62}
  \op U^I(t)\ket{0}\ket{\alpha}  =& \left( \ket{0} - i \alpha g t \ket{1} \right) \ket{\alpha} + \mathcal{O}(g^2)\\
  \begin{split}
    \op U^I(t)\ket{1}\ket{\alpha}  = &\left( \ket{1} - i \frac{\alpha^*}{2} g t \ket{0} \right) \ket{\alpha} \\ 
    &- igt\ket{0}\frac{\partial}{\partial \alpha^*}\ket{\alpha}+ \mathcal{O}(g^2)
  \end{split}
\end{align}
where
\begin{align}
  \label{eq:63}
  \frac{\partial}{\partial \alpha^*}\ket{\alpha} := &
  \frac{\partial}{\partial
    \alpha^*}\left[\exp{\left(-\frac{|\alpha|^2}{2}\right)}
    \sum\limits_{n=0}^\infty \frac{\alpha^n}{\sqrt{n!}}\ket{n}\right]
\end{align}
($\alpha^*$ is the complex conjugate of $\alpha$).


\end{document}